\documentstyle[prl,aps]{revtex}

\begin{document}

\draft \preprint{MSUPHY00.01}
\title{Do semiclassical zero temperature black holes exist?}

\author{Paul R. Anderson \cite{PRA}}

\address{Department of Physics, Wake Forest University,
Winston-Salem, North Carolina 27109}

\author{ William A.\ Hiscock\cite{BH}, Brett E.\ Taylor\cite{BTA}}

\address{Department of Physics, Montana State University, Bozeman,
Montana 59717 }

\date{\today}
\maketitle



\begin{abstract}
The semiclassical Einstein equations are solved to first order in
$\epsilon = \hbar/M^2$ for the case of a Reissner-Nordstr\"{o}m
black hole perturbed by the vacuum stress-energy of quantized
free fields. Massless and massive fields of spin $0$, $1/2$, and
$1$ are considered. We show that in all physically realistic
cases, macroscopic zero temperature black hole solutions do not
exist. Any static zero temperature semiclassical black hole
solutions must then be microscopic and isolated in the space of
solutions; they do not join smoothly onto the classical extreme
Reissner-Nordstr\"{o}m
solution as $\epsilon \rightarrow 0$.
\end{abstract}

Static spherically symmetric zero temperature black holes have
proven to be very interesting and important at the classical,
semiclassical, and quantum levels. Classically the only static
spherically symmetric black hole solution to Einstein's equations
with zero surface gravity (and hence zero temperature) is the
extreme Reissner-Nordstr\"{o}m (ERN) black hole, which possesses
a charge equal in magnitude to its mass. At the quantum level,
the statistical mechanical entropy of zero temperature (extreme)
black holes has been calculated in string
theory~\cite{strominger:1996} and shown to be identical to the
usual Bekenstein-Hawking formula for the thermodynamic entropy.
The usual semiclassical temperature and entropy calculations for
ERN black holes have all been made in the test field
approximation where the effects of quantized fields on the
spacetime geometry are not considered. However, it is well known
that quantum effects alter the spacetime geometry near the event
horizon of a black hole.  In particular they can change its
surface gravity and hence its
temperature~\cite{bardeen:1981,York,HKY,AHWY,THA}.

In this Letter we examine the effects of the semiclassical
backreaction due to the vacuum stress-energy of massless and
massive free quantized fields with spin $0$, $1/2$, and $1$ on a
static Reissner-Nordstr\"{o}m (RN) black hole.
Our focus is on the effects
the fields have on macroscopic black holes, those substantially larger
than the Planck mass.  We are specifically interested in those
macroscopic black hole configurations that may have zero temperature
when semiclassical effects are incorporated.  Such configurations must
be nearly extreme; that is they must have a charge to mass ratio near
unity.  The fields are
assumed to be in the Hartle-Hawking state, which is a thermal state
at the black hole temperature.  At the event horizon the stress-energy
of quantized fields in the Hartle-Hawking state should be of
order $\epsilon = \hbar/M^2$ compared
to the stress-energy of the classical electric field, with $M$
the mass of the black hole.  Thus semiclassical effects
may be handled using perturbation theory.

In all physically realistic cases we find that solutions to the
perturbed semiclassical backreaction equations corresponding to
static spherically symmetric zero temperature black holes do not
exist. In the context of semiclassical gravity with free
quantized fields as the matter source, this means that no
macroscopic zero temperature static black hole solutions exist.
This is a very surprising and general result that may have
significant implications for black hole thermodynamics. If there
are any zero temperature static black hole solutions within the
full semiclassical theory of gravity (not perturbation theory),
then those solutions must be isolated in the space of solutions
from the classical extreme Reissner-Nordstr\"{o}m solution. That
is, they cannot join smoothly onto the ERN solution as
$\hbar/M^2$ approaches zero.

The general static spherically symmetric metric can be written in
the form \cite{MTW}:
\begin{equation}
        ds^2 = - f(r)dt^2 + h(r) dr^2 + r^2 d\Omega^2 \; ,
\label{gen_fh}
\end{equation}
where $d\Omega^2$ is the metric of the two-sphere. The metric can
describe a black hole with an event horizon at $r = r_h$ if
$f(r_h) = 0$.  To avoid having a scalar curvature singularity at
the event horizon it is necessary that $h^{-1}(r_h) = 0$ as well
\cite{sizegap}. The surface gravity of such a black hole is
\begin{equation}
        \kappa = \left(\frac{1}{2}\right)\frac{f'}{\sqrt{f h}}
                \Biggm| _{r = r_h} \; ,
\label{kappa}
\end{equation}
where the prime represents a derivative with respect to $r$ and
the expression is evaluated at the horizon radius, $r_h$. The
temperature is then \cite{waldbook} $T = \kappa/(2 \pi)$.

Since we wish to perturb the spacetime with the vacuum energy of
quantized fields, we begin by considering the general
Reissner-Nordstr\"{o}m metric as the ``bare'' state.  For the RN
metric,
\begin{equation}
    f(r) = h^{-1}(r) = 1 -  \frac{2M}{r} + \frac{Q^2}{r^2} \;,
\label{rn_metric}
\end{equation}
where $Q$ is the electric charge and $M$ is the mass of the black
hole. The outer event horizon is located at
\begin{equation}
        r_{+} = M + \sqrt{M^2-Q^2} \; .
\label{rplus}
\end{equation}
For the ERN black hole $\vert Q \vert = M$.

In semiclassical gravity, the geometry is treated classically
while the matter fields are quantized.  In examining the
semiclassical perturbations of the RN metric caused by the vacuum
energy of quantized fields, we continue to treat the background
electromagnetic field as a classical field. The right hand side
of the semiclassical Einstein equations will then contain both
classical and quantum stress-energy contributions,
\begin{equation}
         G^{\mu}\!_{\nu} = 8 \pi \left[\left(T^{\mu}\!_{\nu}\right)^C
         +  \langle  T^{\mu}\!_{\nu} \rangle  \right] \; .
\label{QEFE2}
\end{equation}
We consider the situation where the black hole is in thermal
equilibrium (whether at zero or nonzero temperature) with the
quantized field; the perturbed geometry then continues to be
static and spherically symmetric. To first order in $\epsilon =
\hbar/M^2$ the general form of the perturbed RN metric may be
written as:
\begin{equation}
        ds^2 = -[1 + 2\epsilon \rho(r) ]
                \left(1 - \frac{2m(r)}{r} + \frac{Q^2}{r^2}\right) dt^2
                + \left(1 - \frac{2m(r)}{r} +
                \frac{Q^2}{r^2}\right)^{-1} dr^2 + r^2 d\Omega ^2 \; ,
\label{pertrn}
\end{equation}
The function $m(r)$ contains both the classical mass and a
first-order quantum perturbation,
\begin{equation}
 m(r) = M[1 + \epsilon \mu (r)] \; .
\label{mofr}
\end{equation}
The metric perturbation functions, $\rho (r)$ and $\mu (r)$, are
determined by solving the semiclassical Einstein equations
expanded to first order in $\epsilon$,
\begin{eqnarray}
        \frac{d \mu}{dr} &=& - \frac{4\pi r^2}{M \epsilon}
                \langle T^t\!_t \rangle  \; ,
\label{mu_eqn} \\
        \frac{d \rho}{d r} &=& \frac{4 \pi r}{\epsilon}
                \left(1 - \frac{2M}{r} + \frac{Q^2}{r^2} \right)^{-1}
                \left[\langle T^r\!_r \rangle - \langle T^t\!_t
                \rangle \right]
                \; .
\label{rho_eqn}
\end{eqnarray}
The right hand side of Eq.\ (\ref{rho_eqn}) is divergent on the
horizon unless $[\langle T^r\!_r \rangle
 - \langle T^t\!_t\rangle]$ vanishes there in the RN case and
unless it and its first radial derivative vanish there in the ERN
case. Conservation of stress-energy implies that so long as the
radial derivative of the stress-energy tensor is finite at the
horizon then there is no divergence in the (nonextreme) RN case.
For the extreme case in two-dimensions, Trivedi has shown
\cite{trivedi} that a divergence of this quantity does occur for
the conformally invariant scalar field. However in
four-dimensions Anderson, Hiscock, and Loranz have shown \cite{AHL} by explicit
numerical computation of the renormalized stress-energy of a
quantized massless scalar field that there is no divergence in
the stress-energy at the horizon. If such a divergence did occur
in four-dimensions for some other field, it would indicate that a
freely falling observer passing through the event horizon would
see an infinite energy density there. The perturbation
approximation would break down in this case, even for ERN black
holes with arbitrarily large masses, and hence would be outside
of the scope of this work.

Assuming the perturbation expansion remains valid, the functions
$\mu(r)$ and $\rho(r)$, obtained by integrating Eqs.\ (\ref
{mu_eqn}-\ref{rho_eqn}), will contain constants of integration.
It is convenient to define them as the values of the metric
perturbations on the unperturbed horizon at $r_+$, so that $\mu(r_+) = C_1$
and $\rho(r_+) = C_2$.  Since we are working in perturbation theory,
the values of these quantities on
the actual horizon are, to leading order, also $C_1$ and $C_2$ respectively.
Then to first order in $\epsilon$ the value
of $m(r)$ at the horizon is 
$m(r_h) = M(1+\epsilon C_1)$.  It is clear that
$C_1$ represents a finite renormalization of the mass $M$ of the
black hole. As in previous work\cite{AHWY,THA}, we hereafter
denote the renormalized perturbed mass at the horizon,
 $m(r_h) = M(1+\epsilon C_1)$, by $M_R$. 
The quantity $(1-2m(r_h)/r_h +
Q^2/r_h^2)$ then vanishes at $r_h = r_+$, where now $r_+ = M_R +
(M_R^2-Q^2)^{1/2}$ . Thus this renormalization causes the
perturbed horizon to be located at the same radius $r_{+}$ (as a
function of the physical, renormalized mass $M_R$ and the charge $Q$) as
the classical horizon.

To decide whether a semiclassically perturbed black hole has zero
temperature, we must calculate the surface gravity of the
perturbed metric to first order in $\epsilon$. Applying Eq.\
(\ref{kappa}) to the metric of Eq.\ (\ref{pertrn}) and using
Eqs.\ (\ref{mu_eqn}) and (\ref{rho_eqn}) to simplify the result
gives
\begin{equation}
        \kappa = \frac{\sqrt{M_R^2-Q^2}}{r_{+}^2}\left( 1 +
        \epsilon C_2 \right) + 4 \pi r_+ \langle T^t\!_t \rangle
        \! \! \bigm| _{r = r_+}\; .
\label{eq:kappa_rn}
\end{equation}

Now consider semiclassical black holes that, at first order in
$\epsilon$, have precisely zero temperature. Such black holes are
legitimate solutions within the context of perturbation theory
only if they maintain zero temperature as $\epsilon$ is reduced
to zero. From Eq.\ (\ref{eq:kappa_rn}) it is seen that for the
surface gravity, $\kappa$, to be zero, the classical surface
gravity of the ``bare'' black hole,
\begin{equation}
       \kappa_0 =  \frac{\sqrt{M_R^2-Q^2}}{r_{+}^2} \; ,
\label{kappa0}
\end{equation}
must be at most of order $\epsilon$.  Thus the term in Eq.\
(\ref{eq:kappa_rn}) involving the (unknown) integration constant
$C_2$ will be at least of order $\epsilon^2$, and hence may be
discarded in this case. The total surface gravity of the
semiclassical solution at first order then involves two terms:
the classical surface gravity, which is always nonnegative, and a
term proportional to $\langle T^t\!_t \rangle$.  To have a
semiclassically perturbed zero temperature black hole, it is then
necessary that $\langle T^t\!_t \rangle$ be nonpositive at the
horizon. This implies that the vacuum energy density at the event
horizon must be nonnegative. If the vacuum energy density
is negative at the event horizon (and therefore the
weak energy condition \cite{waldbook} is violated there), then quantum effects will prevent
a zero temperature semiclassical perturbed black hole from existing.

The calculation of the expectation value of the stress-energy of
a quantized field in a curved spacetime is a very difficult
exercise.  However, the problem is simplified in the present case
by our focus on zero temperature solutions. Since the classical
``bare'' solution must have a surface gravity that is of order
$\epsilon$ or less, we can simply consider the vacuum
stress-energy for the ERN spacetime. While the actual bare
spacetime may be slightly non-extreme (to order $\epsilon$), the
differences between the vacuum stress-energy tensor of the
extreme spacetime and the bare spacetime will be of order
$\epsilon^2$, and may be ignored.

The ERN spacetime is asymptotically congruent to the conformally
flat Robinson-Bertotti spacetime as one approaches the event
horizon at $r=M_R$ \cite{Bertotti,Robinson,Carter}. The vacuum
stress-energy of a quantized field should similarly
asymptotically approach the Robinson-Bertotti values as one
approaches the event horizon of the ERN spacetime. This has been
confirmed numerically for the scalar field using point splitting
renormalization\cite{AHL}. For conformally invariant (hence,
massless) quantized fields, the vacuum stress-energy in the
Robinson-Bertotti spacetime may be obtained using the results of
Brown and Cassidy \cite{BrownCassidy} and Bunch\cite{Bunch}. It
is\cite{AHL}
\begin{equation}
        \langle T^\mu\!_\nu \rangle = {b(s) \over 2880 \pi^2 M^4}
        \delta^\mu\!_\nu \; ,
\label{RBTmn}
\end{equation}
with $b(s) = 1, \frac{11}{2}, 62$ for scalar, spinor, and vector
fields respectively. Since  $\langle T^t\!_t \rangle$ is positive
for all three of these cases, the vacuum energy density is
negative in all these cases on the ERN horizon, and hence there
are no zero temperature linearly perturbed RN black holes
associated with conformally invariant quantized fields.

Next let us consider the massless quantized scalar field with
arbitrary curvature coupling, $\xi$ (the scalar field is
conformally invariant only if $\xi = 1/6$). In this case, the
vacuum stress-energy tensor has been numerically computed using
point splitting renormalization for the ERN black hole
spacetime\cite{AHL}.
The vacuum stress-energy depends on $\xi$ in a linear fashion,
and may be divided into conformal and nonconformal pieces:
\begin{equation}
        \langle T^\mu\!_\nu \rangle = C^\mu\!_\nu + \left(\xi
        - \frac{1}{6}\right)D^\mu\!_\nu \; .
\label{scalarTmn}
\end{equation}
Anderson, Hiscock, and Loranz \cite{AHL} found that $
C^\mu\!_\nu$ approaches the Robinson-Bertotti values as $r
\rightarrow M$, and that all components of $D^\mu\!_\nu$ approach
zero in that limit. Hence, at the horizon of an ERN black hole,
the vacuum stress-energy tensor of a quantized scalar field is
independent of the curvature coupling, and is equal to the
Robinson-Bertotti value. Therefore, there are no zero temperature
linearly perturbed RN black holes associated with massless
quantized scalar fields for any value of the curvature coupling.

We also wish to consider quantized massive fields in the ERN
black hole spacetime. The vacuum stress-energy of quantized
massive fields in the RN spacetime has  been numerically computed
using point splitting renormalization in the case of scalar
fields, by Anderson, Hiscock, and Samuel \cite{AHS}. They also
developed the DeWitt-Schwinger approximation $\langle T^\mu\!_\nu
\rangle_{DS}$ for the stress-energy of the massive scalar field,
and found that the exact values of the stress-energy components
were well approximated when the black hole mass $M$ and field
mass $m$ satisfy $Mm > 2$ (it does not matter here whether $M$ is
the bare or renormalized black hole mass; any resulting
difference will be higher order in $\epsilon$). As the field mass
is increased, the DeWitt-Schwinger approximation rapidly becomes
more accurate. The DeWitt-Schwinger approximate value for the
vacuum energy density of a massive scalar field, evaluated at the
event horizon of an ERN black hole is
\begin{equation}
        \langle T^t\!_t \rangle_{DS} \Biggm| _{r = M} =
        {\epsilon (5 - 14 \xi) \over 10080 \pi^2 M^4 m^2} \; .
\label{TttDSsc}
\end{equation}
Zero temperature perturbed solutions will only be possible if
$\langle T^t\!_t \rangle$ is negative. Examination of Eq.\
(\ref{TttDSsc}) shows that will only be possible if
        $\xi \geq {5 \over 14}$,
a range that excludes the cases of greatest physical interest,
namely the minimally ($\xi = 0$) and conformally ($\xi = 1/6$)
coupled fields. A thorough study of RN black holes (with
arbitrary charge) perturbed by a quantized massive scalar field
has been presented elsewhere \cite{THA}.

The DeWitt-Schwinger approximation has recently been extended to
the case of massive spinor and vector fields in the RN black
hole spacetime by Matyjasek\cite{Maty}. The accuracy of the
DeWitt-Schwinger approximation is unknown in this case, as no
direct calculation of the exact value of $\langle T^\mu\!_\nu
\rangle$ has been performed for these fields in the RN spacetime.
For the spinor field around an ERN black hole, Matyjasek finds
\begin{equation}
        \langle T^t\!_t \rangle_{DS} \Biggm| _{r = M} =
        {37 \epsilon \over 40320 \pi^2 M^4 m^2} \; ,
\label{TmnDSsp}
\end{equation}
while for the vector field, he obtains
\begin{equation}
        \langle T^t\!_t \rangle_{DS} \Biggm| _{r = M} =
        {19 \epsilon \over 3360 \pi^2 M^4 m^2} \; .
\label{TmnDSv}
\end{equation}
Since both of these values for $\langle T^t\!_t \rangle$ are
manifestly positive, it appears that perturbations of an ERN
black hole caused by quantized massive spinor or vector fields
cannot yield a zero temperature solution.

Finally we note that in general there are higher derivative terms
in the semiclassical backreaction equations which come from terms
in the gravitational action that are quadratic in the curvature.
These terms can be taken into account perturbatively by putting
them on the right hand side of the equations and evaluating them
in the background geometry \cite{CLA}.  The effective
stress-energy tensor for these terms vanishes at the event
horizon in the ERN geometry. Thus these terms cannot cancel the
effects of the negative energy densities due to the quantized
fields.

Our results imply that if static zero temperature semiclassical
black hole solutions do exist, they must not smoothly join onto
the classical zero temperature ERN solution as $\epsilon =
\hbar/M^2 \rightarrow 0$. This suggests that any such solutions
are truly microscopic, with masses within a few orders of the
Planck mass. Whether such small zero temperature black hole
solutions exist remains an open question.

One implication of the nonexistence of macroscopic zero
temperature black hole solutions is that, for fixed mass M, there
is a minimum temperature that any static spherically symmetric
semiclassical black hole can have, namely (from
Eq.(\ref{eq:kappa_rn})), $ T = 2 r_+ \langle T^t\!_t \rangle \!
\! \bigm| _{r = r_+}$ . Thus it is not only impossible to build a
macroscopic zero temperature black hole~\cite{BCH}, it is
impossible to build one that is arbitrarily close to zero
temperature.  This is a reformulation of one version of the third
law of black hole mechanics~\cite{waldbook}.

This work was supported in part by National Science Foundation
Grant No. PHY-9734834 at Montana State University and No.
PHY-9800971 at Wake Forest University.



\begin{references}

\bibitem[*]{PRA}
electronic mail address: anderson@wfu.edu

\bibitem[\dagger]{BH}
electronic mail address:  hiscock@physics.montana.edu

\bibitem[\ddagger]{BTA}
present address: Department of Chemistry and Physics,
Radford University, Radford, VA 24142\\
%
electronic mail address:  brett@peloton.runet.edu

\bibitem{strominger:1996} A.\ Strominger and C.\ Vafa, Phys. Lett. {\bf
B379}, 99 (1996).

\bibitem{bardeen:1981} J.\ M.\ Bardeen, Phys.\ Rev.\ Lett.\ {\bf 46},
382 (1981).

\bibitem{York} J.\ W.\ York,  Jr., Phys.\ Rev.\ D {\bf 31}, 775 (1985).

\bibitem{HKY} D.\ Hochberg, T.\ W.\ Kephart, and J.\ W.\ York, Jr.,
        Phys.\ Rev.\ D {\bf 48}, 479 (1993).

\bibitem{AHWY}P.\ R.\ Anderson, W.\ A.\ Hiscock, J.\ Whitesell, and
        J.\ W.\ York Jr., Phys.\ Rev.\ D {\bf 50}, 6427 (1994).


\bibitem{THA} B.\ E.\ Taylor, W.\ A.\ Hiscock, and P.\ R.\ Anderson,
        Phys.\ Rev.\ D.\ {\bf 61}, 84021 (2000).

\bibitem{MTW}Throughout we use units such that $\hbar = c = G = k_B =
1$.  Our
        conventions are those of C.\ W.\ Misner, K.\ S.\ Thorne, and J.\
A.\ Wheeler,
        {\it Gravitation } (Freeman, San Francisco, 1973).

\bibitem{sizegap} P.\ R.\ Anderson and C.\ D.\ Mull, Phys.\ Rev.\ D.\
{\bf 59},
         044007 (1999).

\bibitem{waldbook} R.\ M.\ Wald, {\it General Relativity} (University of
Chicago
         Press, Chicago, 1984).

\bibitem{trivedi} S.\ P.\ Trivedi, Phys.\ Rev.\ D{\bf 47}, 4233 (1993).

\bibitem{AHL} P.\ R.\ Anderson, W.\ A.\ Hiscock, and D.\ J.\ Loranz,
        Phys.\ Rev.\ Lett.\ {\bf 74}, 4365 (1995).

\bibitem{Bertotti} B.\ Bertotti, Phys.\ Rev.\ {\bf 116}, 1331 (1959).


\bibitem{Robinson} I.\ Robinson, Bull.\ Acad.\ Pol.\ Sci.\ {\bf 7},
        351 (1959).


\bibitem{Carter} B.\ Carter, {\it Black Holes}, edited by C.\ DeWitt
        and B.\ S.\ DeWitt (Gordon and Breach, New York, 1973).


\bibitem{BrownCassidy} L.\ S.\ Brown and J.\ P.\ Cassidy, Phys.\ Rev.\
        D {\bf 15}, 2810 (1977).

\bibitem{Bunch} T.\ S.\ Bunch, J.\ Phys.\ A {\bf 12}, 517 (1979).


\bibitem{AHS} P.\ R.\ Anderson, W.\ A.\ Hiscock, and D.\ A.\ Samuel,
Phys.\
        Rev.\ D {\bf 51}, 4337 (1995).

\bibitem{Maty} J.\ Matyjasek, Phys.\ Rev.\ D {\bf 61}, 124019 (2000).

\bibitem{CLA} M.\ Campanelli, C.\ O.\ Lousto, and J.\ Audretsch, Phys.
Rev. D{\bf 49},
5188 (1994); Phys. Rev. D{\bf 51}, 6810 (1995).

\bibitem{BCH} J.\ M.\ Bardeen, B.\ Carter, and S.\ W.\ Hawking, Commun.\ Math.\ 
Phys. {\bf 31}, 161 (1973).


\end{references}
\end{document}